# Analysis of large effective electric fields of weakly polar molecules for electron electric dipole moment searches

A. Sunaga, M. Abe*, and M. Hada

*Tokyo Metropolitan University, 1-1, Minami-Osawa, Hachioji-city, Tokyo 192-0397, Japan*

B. P. Das

*Department of Physics and International Education and Research Center of Science*
*Tokyo Institute of Technology, 2-12-1-H86 Ookayama, Meguro-ku, Tokyo152-8550, Japan*

The electric dipole moment of an electron (eEDM) is one of the sensitive probes of physics beyond the standard model. The possible existence of the eEDM gives rise to an experimentally observed energy shift, which is proportional to the effective electric field ($E_{\text{eff}}$) of a target molecule. Hence, an analysis of the quantities that enhance $E_{\text{eff}}$ is necessary to identify suitable molecules for eEDM searches. In the context of such searches, it is generally believed that a molecule with larger electric polarization also has a larger value of $E_{\text{eff}}$. However, our Dirac-Fock and relativistic coupled-cluster singles and doubles calculations show that the hydrides of Yb and Hg have larger $E_{\text{eff}}$ than those of fluorides, even though their polarizations are smaller. This is due to significant mixing of valence $s$ and $p$ orbitals of the heavy atom in the molecules. This mixing has been attributed to the energy differences of the valence atomic orbitals and the overlap of the two atomic orbitals based on the orbital interaction theory.

## I. INTRODUCTION

The electric dipole moment (EDM) of the electron (eEDM) is an important probe of physics beyond the Standard Model of particle interactions [1]. It can be obtained by combining the results of the measured values of the energy shifts in an atom or a molecule with the calculated values of the effective electric field ($E_{\text{eff}}$), which can be interpreted as the net electric field experienced by an electron in an atom or a molecule. The first experiment to detect the eEDM was performed on cesium atom [2]. However, from reasons first put forward by Sandars [3], it became evident that polar molecules, such as halides containing a heavy atom, would have relatively large $E_{\text{eff}}$, and would therefore be more suitable for such an experiment [4,5] than a heavy atom.

The enhancement of $E_{\text{eff}}$ is due to the mixing of the valence $s$ orbital with $p$ orbitals ($s$-$p$ mixing). In the atomic case, an external electric field is needed for the $s$-$p$ mixing, but this mixing is extremely small because of the limit of the external field attainable in a laboratory. In contrast, in a heavy polar molecule containing a halogen atom, a valence electron in the heavy atom moves to the halogen atom. The electron is localized in the halogen and it produces an electric field which is much larger than the external electric field in an atomic experiment. As a result, the heavy atomic ion experiences an electric field which comes from the electron. Therefore, in the polar molecule, the valence orbitals of $s$ and $p$ mix much more than in an atom and the mixing causes a larger $E_{\text{eff}}$ [4].

In addition, $E_{\text{eff}}$ of a nonpolar molecule would appear to be small, because the bond between atoms is not ionic but covalent. In this case, the valence electron would be delocalized in the whole molecule, and it would not feel the strong electric field created by the heavy atomic nucleus.

There are many previous theoretical studies on halides containing a heavy atom [6,7] based on the points mentioned in the earlier two paragraphs . In the past five years, YbF [8] and ThO [9], which are *polar molecules*, have led to new upper limits of the eEDM. Recently, RaF [10] and HgX [11] (X is a halogen) have been proposed as new candidates for the search of eEDM. It is a commonly held idea that as the molecules become more polarizable their $E_{\text{eff}}$ becomes larger. This is even mentioned in a recent article in Physics Today [12].

However, our calculations in this paper show that hydrides of ytterbium and mercury (YbH and HgH) have larger $E_{\text{eff}}$ than those of fluorides (YbF and HgF) at both the Dirac-Fock (DF) and relativistic coupled cluster singles and doubles (RCCSD) levels. On the other hand, the permanent dipole moments (PDMs) of the hydrides are much smaller than those of the fluorides, which indicates that the electric polarizations of the hydrides are relatively smaller. The results we have obtained for $E_{\text{eff}}$ and PDMs are not what was expected from the previous works [4,5,12].

In order to explain the reason for our anomalous findings, we analyzed our calculated DF orbitals using the Mulliken population (MP) analysis [13]. We found that the tendency of the polarization of all electrons is same to the tendency of PDM. Both of them show the fluorides have larger polarization than the hydrides. However, the single occupied molecular orbital (SOMO), which mainly contributes to $E_{\text{eff}}$, are almost localized in the heavy atom in all the four molecules. In addition, in the hydrides, $p$ orbitals in the heavy atom contribute to the SOMO more than in the fluorides, and hence, the hydrides have larger $E_{\text{eff}}$ than the fluorides. The larger contribution of $p$ would originate from the smaller energy gap and larger overlap integral between the valence orbitals of each atom in the molecules, based on the orbital interaction theory [14, 15].

## II. THEORY

The molecular properties were calculated at the Dirac-Fock (DF) and the relativistic coupled-cluster single and double (RCCSD) levels. The exact wavefunction $|\Psi\rangle$, of a quantum many-body system can be written in the framework of the coupled-cluster method as

$$|\Psi\rangle = e^{\hat{T}}|\Phi_0\rangle, \quad (1)$$

where $T$ is the cluster operator and $|\Phi_0\rangle$ is the reference wavefunction. In our calculations, it was chosen to be the DF wavefunction corresponding to the ground state of the molecule, which was a Slater determinant composed of single particle four-component spinors.

The cluster operator $\hat{T}$ is defined as

$$\hat{T} = \sum_k^{N_e} \hat{T}_k. \quad (2)$$

$$\hat{T}_1 = \sum_{i,a} t_i^a \hat{a}_i \hat{a}_a^\dagger, \quad \hat{T}_2 = \sum_{i>j, a>b} t_{ij}^{ab} \hat{a}_i \hat{a}_j \hat{a}_a^\dagger \hat{a}_b^\dagger, \quad \hat{T}_3 = \cdots, \quad (3)$$

where $N_e$ is the number of electrons in the molecule. $i$ and $j$ ($a$ and $b$) are orbital indices which are occupied (unoccupied) in $|\Phi_0\rangle$. $t_i^a$ and $t_{ij}^{ab}$ represent cluster amplitudes and $\hat{a}$ and $\hat{a}^\dagger$ are the annihilation and the creation operators, respectively. In this work, we have used the CCSD approximation which is expressed as

$$\hat{T} \approx \hat{T}_1 + \hat{T}_2. \quad (4)$$

The CCSD approximation contains certain higher order terms that are not present in the configuration interaction singles and doubles (CISD) approximation. The reason for this is the exponential nature of the coupled-cluster wavefunction as given in Eq. (1), which gives rise to non-linear terms.

The eEDM interaction in the molecule used by us is given by the following effective operator [16,17]

$$\hat{H}_{\text{eEDM}} = -2d_e ic \sum_j^{N_e} \beta \gamma_5 p_j^2. \quad (5)$$

Here, $d_e$ is the value of the electron electric dipole moment, $i$ is the imaginary unit, and $c$ is the speed of light. $\beta$ and $\gamma_5$ are the four-component Dirac matrices and $p$ is the momentum operator. The above expression consists of the effect of the electric field due to both the nuclei and the electrons in the molecule.

The effective electric field in the molecule can be expressed as

$$E_{\text{eff}} = \left\langle \Psi \left| \frac{\hat{H}_{\text{eEDM}}}{d_e} \right| \Psi \right\rangle. \quad (6)$$

The permanent dipole moment (PDM) of the molecule was evaluated by using the expression

$$\text{DM} = -\left\langle \Psi \left| \sum_i^{N_e} \mathbf{r}_i \right| \Psi \right\rangle + \sum_A^{N_A} Z_A \mathbf{R}_A, \quad (7)$$

where $N_A$ and $N_e$ are the total number of nuclei and electrons, respectively. $\mathbf{R}$ and $\mathbf{r}$ refer to the position vectors of electrons and nuclei, and $Z$ is the nuclear charge.

## III. COMPUTATIONAL DETAILS

Our calculations were based on the RCCSD approximation by combing the UTCHEM [19] and DIRAC08 [20] codes. $E_{\text{eff}}$ was computed by the modified UTCHEM [21]. UTCHEM was employed for the generation of the Dirac-Fock orbitals and the molecular orbital integral transformation using the Dirac-Coulomb Hamiltonian. DIRAC08 was used for the evaluation of the RCCSD amplitudes.

In this work, we used primitive Gaussian basis sets. For Yb and Hg atoms, Dyall double-zeta (DZ) and quadrupole-zeta (QZ) basis sets [22] were used. For H and F atoms, Watanabe's four-component basis sets [23] were used in all the calculations. We added some diffuse and polarization functions to each above basis set. These functions were taken from the Dyall basis sets for Hg and Yb [22], and Sapporo basis sets [24] for H, F and Yb. The basis sets used for Yb and F were same as that used in the previous work [17]. In the calculation of HgF at the QZ level, some of the diffuse Sapporo basis functions, $1d1f$, were removed for F atom to avoid a convergence problem. The size of the basis sets employed in this work are shown in Table I and the used exponential parameters are in the supplementary file [25]. The cutoff values for the energies of the virtual molecular orbitals for our calculations were 80 a.u. In our CCSD calculations, all the electrons in the molecules could be excited. The experimental bond lengths were used for the calculations of YbH, YbF and HgH molecules. The used bond length of HgF was optimized by Knecht *et al.* using four-component Fock-space CCSD level. They are 2.0526, 2.0161, 1.7662 and 2.00686 for YbH [26], YbF [27], HgH [26] and HgF [28], respectively in Angstrom.

The cluster amplitudes, $t_i^a$ and $t_{ij}^{ab}$ were evaluated using the RCCSD method, but the expectation values were calculated by using only the linear terms in the RCCSD wavefunction, written as [17,29,30]

$$\langle \Phi_0 | (1 + T_1^+ + T_2^+) \hat{O}_N (1 + \hat{T}_1 + \hat{T}_2) | \Phi_0 \rangle_C$$
$$+ \langle \Phi_0 | \hat{O} | \Phi_0 \rangle. \quad (8)$$

The most important contributions are captured in this treatment with feasible computational costs. $E_{\text{eff}}$ and PDM of the molecules were estimated by using the above equation.

## IV. RESULTS AND DISCUSSION

Table II shows our previous [17] and present results at the DF and CCSD levels using the Dyall-DZ and QZ basis sets for all the molecules that we have considered. The latest results for HgH based on the relativistic CCSD z-vector method with QZ basis sets has been reported by Sasmal *et al.* [18] and they are also given in this table. Our HgH result agrees with that of Sasmal *et al.* to about 4%. The relatively small disagreement between the two results is due to the difference in the basis sets, the cutoff energy value for CCSD, and the approximation made in our property calculations.

For Yb and Hg systems, the hydrides have larger $E_{\text{eff}}$ than the fluorides at both the DF and CCSD levels, but for PDM the trend is just the opposite; i.e. the values of the hydrides are smaller than those of fluorides. This is contrary to conventional wisdom: "molecules with smaller polarization would have smaller $E_{\text{eff}}$". The differences in the values of $E_{\text{eff}}$ between the hydrides and the fluorides increase at the CCSD level compared to the DF level. The difference in $E_{\text{eff}}$ between HgH and HgF at the CCSD and QZ basis level is about 3.7 %, which is smaller than the estimated computational error (6~8%) obtained from the experimental comparison in our previous work [17,20]. In contrast, the difference in $E_{\text{eff}}$ between YbH and YbF at the same level is about 26.2 %, which is substantially larger than the above estimated error. Therefore we conclude with certainty that the hydrides have larger $E_{\text{eff}}$ than the fluorides at least in Yb systems.

Electron correlation increases $E_{\text{eff}}$ for all the molecules we have considered, but it increases the PDM for the Yb systems and decreases that for the Hg systems. The PDM values of YbF, HgF, and YbH are relatively large (3.59 ~2.93), but the PDM of HgH is quite small (0.15 D), compared to the other three molecules. This is qualitatively explained from the difference in the electronegativities between the atoms that make up these molecules. The values of the Allred-Rochow electronegativities of hydrogen, fluorine, ytterbium, and mercury are 2.20, 4.10, 1.06 and 1.44, respectively [31,32]. The largest difference in the electronegativities is between fluorine (4.10) and ytterbium (1.06) and the calculated PDM in YbF is also the largest (3.59D). Similarly, the order of the differences between the electronegativities of the two atoms is the same as the order of PDMs of the molecules they make up in our calculations.

The relationship between PDM and $E_{\text{eff}}$ is counterintuitive in the case of HgH. This molecule has the largest $E_{\text{eff}}$ in spite of possessing the smallest PDM. As mentioned in the INTRODUCTION it is common to think that a molecule with small polarization would not have a large $E_{\text{eff}}$. The reasons are that (i) the bond is not ionic and (ii) the electric field created in the molecule is smaller than the polar molecule, and former provides smaller *s-p* mixing.

To clarify the reason of this discrepancy, we performed the Mulliken population (MP) analysis [13] by modifying UTCHEM. MP indicates the number of electrons belong to each atomic orbital in a molecule. Table III shows the total MP of all the occupied orbitals at the DF level. The contributions of *g*, *h* and *i* orbitals of the heavy atoms and those of *d* and *f* orbitals of the light atoms are so small that we omit them in the table. From Table III, the number of *s* electrons of the heavy atoms (i.e. Yb or Hg) for all the molecules decreases from 12, which is the number of *s* electrons in the neutral Yb or Hg atom. In contrast, the number of *p* electrons of the heavy atoms increases from 24, the number of *p* electrons in the neutral atoms. (The reason for this increase would be *s-p* mixing in SOMO, which will be explained later.) For the light atoms (i.e. H or F), the number of *s* electrons increases in hydrogen and the number of *p* electrons increases in fluorine. It can be seen from the total MP of each atom in Table III that hydrogen or fluorine becomes highly anionic in YbH, YbF, and HgF. The charge differences between the neutral atoms are 0.59, 0.81, and 0.54 in YbH, YbF, HgF, respectively. In contrast, the hydrogen atom in HgH is much less anionic: The deviation from the neutral hydrogen is only 0.15. These results are consistent with the trend of our calculated PDMs (i.e. small

PDM in HgH) and our expectations on this basis of the electronegativities of the atoms.

In contrast to PDM, which depends on all the occupied orbitals, $E_{eff}$ at the DF level only depends on the SOMO. This is because the contributions of the closed shell orbitals cancel out in the calculation of $E_{eff}$ [17]. Table IV shows the MPs only for the SOMO contributions. In YbH, YbF and HgF, the total MPs of the light atoms in SOMOs are less than 0.01. In HgH, the total MP of hydrogen in SOMO is 0.17, which is a little bigger than the ones in the other molecules. However, the SOMO in HgH is still almost localized in Hg, and hence, it can have a large $E_{eff}$.

In spite of the smaller MPs of the heavy atoms in SOMOs, hydrides have larger $E_{eff}$ than the fluorides. The reason for this could be explained by the contribution of the virtual $6p$ orbitals of the heavy atoms to SOMO. In the four molecules, the valence $6s$ orbitals of the heavy atoms mainly contribute to the SOMOs, but the virtual $6p$ orbitals can also contribute to them. It is well known that the mixing of $s$ and $p$ orbitals ($s$-$p$ mixing) in SOMO is important for a non-zero value of $E_{eff}$, and the large $s$-$p$ mixing in a heavy atom can increase $E_{eff}$ [33]. Hence the larger $s$–$p$ mixing (i.e. the larger contribution of $p$ orbitals), which is shown in table IV is related to the larger values of $E_{eff}$ in the hydrides.

We now turn to why the hydrides have larger $s$-$p$ mixings than the fluorides. In molecules, the $s$ and $p$ orbitals belonging to the same atom cannot strongly interact with each other directly. This is because the overlap of the two orbitals is zero due to the orthogonality of atomic orbitals [14]. (Strictly speaking, they can interact directly because of the nuclear potential of the light atom in the molecular Hamiltonian, but the interaction would be small.)

The $s$-$p$ mixings in the molecules can be interpreted by the orbital interaction theory [14, 15]. In this theory, two atomic orbitals can strongly interact with each other and form a molecular orbital, if the atomic orbital (AO) energy difference is small and the overlap integral of the two orbitals is large. In this theory, the $6s$ and $6p$ orbitals are mixed by the following two steps. (i) The $6s$ of the heavy atom interacts with the valence $s$ or $p$ orbital of the light atom. (ii) The valence orbital of the light atom interacts with the $6p$ virtual orbital of the heavy atom. In this two-step process, $6s$ and $6p$ can mix in SOMO indirectly and intermediate mixing of the valence orbital of the light atom is important for $6s$-$6p$ mixing. In the following paragraph, we consider only the effect of step (i) for simplicity.

Table V shows the AO energy differences and overlap integrals between $6s$ orbital of the heavy atoms and the valence orbital of the light atoms for the four molecules. The orbital energies were obtained from atomic DF calculations using the GRASP2K code [34]. The overlap integrals were obtained by using the contracted Dyall QZ basis sets for the heavy atoms and the contracted Watanabe basis sets for the light atoms. These energy differences show negative correlation with the SOMO-MPs of $p$ orbitals of the heavy atoms. Besides, these overlap integrals show positive correlation with the MPs. From these results, the AO energy differences and overlap integrals would be related to the $s$-$p$ mixing, as the orbital interaction theory suggests. It is also possible to explain why $s$-$p$ mixing in HgF is larger than in YbF because of the smaller energy difference in HgF. The energy diagram for the atomic and molecular orbital energies of the four molecules is shown in figure 1 for ease of understanding.

The smaller energy difference and the larger overlap integral that we have referred to earlier, however, would not be proportion to $E_{eff}$ and not always increase it. There are two reasons for this. Firstly, the key point for a large value of $E_{eff}$ is the *balance* between $s$ and $p$ orbitals. If the contribution of the $p$ orbitals of the heavy atom increases, then that of the $s$ orbitals necessarily decreases. For example, the corresponding contribution of the $s$ orbitals of Hg in HgH is smaller than it is in HgF, because the contribution of the $p$ orbitals of Hg in HgH is larger than it is in the case of HgF. The net result is that the $E_{eff}$ for HgH is a little larger than for HgF, although for YbH it is relatively larger than in YbF. Secondly, when the energy difference is smaller, not only the $6p$ orbital of the heavy atom but also the valence orbital of the light atom largely contributes to SOMO. Since the total MP in SOMO is always one, the large contribution of the light atom leads to a decrease in the contribution of the heavy atom in SOMO, as the MP for HgH indicates. In YbH, YbF and HgF, the total MPs of the heavy atoms in SOMO are more than 0.9, but that of HgH is 0.83. That smaller MP of the heavy atom would be the reason why HgH has a slightly larger $E_{eff}$ than HgF. If the energy difference were very closed to zero, the valence orbital of the light atom would greatly contribute to SOMO. As a results, the SOMO electron would not localized to the heavy atom and $E_{eff}$ would greatly decrease.

The analysis we have presented above is based on the orbital interaction theory. It can explain why the hydrides have larger $s$-$p$ mixings in SOMO than the fluorides and it is the reason why the hydrides have a comparatively larger $E_{eff}$. These explanations are not possible by the conventional idea based on the electric polarization of molecules. In addition, our idea would help to search for new candidate molecules with large $E_{\text{eff}}$. Based on the Koopmans's theorem, we may qualitatively estimate AO energy differences from experimental ionization energies (IE). Hence, we can identify some molecules which would have large $s$-$p$ mixings in SOMO, and that would be helpful to suggest new candidate molecules with large $E_{\text{eff}}$.

## IV. CONCLUSION

We find that the hydrides (YbH, HgH) have larger values of $E_{\text{eff}}$ than the fluorides (YbF, HgF) and explained the reason for this using the Mulliken population analysis and the orbital interaction theory. The conventional concept that molecules with small polarizations have small $s$-$p$ mixing and $E_{\text{eff}}$, could not explain the trend of $E_{\text{eff}}$ for the fluorides and hydrides. Instead, we consider the mixing of $s$ and $p$ orbitals in SOMO would be derived from the energy difference and overlap integral of the valence orbitals of the two atoms based on the orbital interaction theory. It has been argued on the basis of orbital interaction theory that large $s$-$p$ mixing in SOMO is essential to search for new candidates for eEDM experiments. For molecules, which can be described by a single reference method, we would adopt the approach presented in this paper to find other molecules with large $E_{\text{eff}}$. Since the atomic valence energy is equal to the ionization energy on the basis of the Koopman theorem, the atomic orbital energy differences can be qualitatively estimated from experimental ionization energies. This can facilitate the search of new candidates with large $E_{\text{eff}}$ for the detection of the eEDM.

## ACKNOWLEDGEMENT

We would like to thank Professor Y. Imamura and Professor H. Kanamori, for valuable discussion. This study was supported by the Core Research for Evolutional Science and Technology (CREST) program from the Japan Science and Technology (JST) Agency.

[1] M. Pospelov, A. Ritz, Ann. Phys. (NY) **318**, 119 (2005); T. Fukuyama, Int. J. Mod. Phys. A **27**, 1230015 (2012); T. Ibrahim, A. Itani, and P. Nath, Phys. Rev. D **91**, 095017 (2015).
[2] P. G. H. Sandars and E. Lipworth, Phys. Rev. Lett. **13**, 718 (1964).
[3] P. G. H. Sandars, Phys. Rev. Lett. **19**, 1396 (1967).
[4] O. P. Sushkov, V. V. Flambaum, and I. B. Khriplovich, Zh. Eksp. Teor. Fiz 87, 1521 (1984).
[5] M. G. Kozlov and V. F. Ezhov, Phys. Rev. A **49**, 4502 (1994)
[6] A. V. Titov, N. S. Mosyagin, A. N. Petrov, T. A. Isaev and D. P. DeMille, in *Recent Advances in the Theory of Chemical and Physical Systems*, (Springer, Berlin, 2006), p.253.
[7] V. A. Dzuba, V. V. Flambaum, and C. Harabati, Phys. Rev. A **84**, 052108 (2011)
[8] J. J. Hudson, D. M. Kara, I. J. Smallman, B. E. Sauer, M. R. Tarbutt and E. A. Hinds, Nature (London) **473**, 493 (2011).
[9] The ACME Collaboration, J. Baron *et al.*, Science **343**, 269 (2014).
[10] T. A. Isaev, S. Hoekstra, and R. Berger, Phys. Rev. A **82**, 052521 (2010).
[11] V. S. Prasannaa, A. C. Vutha, M. Abe, and B. P. Das, Phys. Rev. Lett. **114** 183001 (2015).
[12] D. DeMille, Phys. Today **68**(12), 34 (2015).
[13] R. S. Mulliken, J. Chem. Phys. **23**, 1833 (1955).
[14] L. Libit and R. Hoffmann, J. Am. Chem. Soc, **96**, 1370 (1974).
[15] S. Inagaki, H. Fujimoto and K. Fukui, J. Am. Chem. Soc, **98**, 4054 (1974).
[16] B. P. Das, in *Aspects of Many-Body Effects in Molecules and Extended Systems*, edited by D. Mukherjee (Springer, Berlin, 1989), p. 411.
[17] A. Sunaga, M. Abe, B. P. Das and M. Hada, Phys Rev. A **93**, 042502 (2016).
[18] S. Sasmal, H. Pathak, M. Nayak, N. Vaval and S. Pal, J. Chem. Phys **144**, 124307 (2016).
[19] T. Yanai et al., in UTCHEM—A Program for ab initio Quantum Chemistry, edited by G. Goos, J. Hartmanis, and J. van Leeuwen, Lecture Notes in Computer Science Vol. 2660 (Springer, Berlin, 2003), p. 84; T. Yanai, T. Nakajima, Y. Ishikawa, and K. Hirao, J. Chem. Phys. **114,** 6526 (2001); **116**, 10122 (2002).
[20] DIRAC, a relativistic ab initio electronic structure program, Release DIRAC08 (2008), written by L. Visscher, H. J. Aa. Jensen, and T. Saue, with new contributions from R. Bast, S. Dubillard, K. G. Dyall, U. Ekström, E. Eliav, T. Fleig, A. S. P. Gomes, T. U. Helgaker, J. Henriksson, M. Iliaš, Ch. R. Jacob, S. Knecht, P. Norman, J. Olsen, M. Pernpointner, K. Ruud, P. Sałek, and J. Sikkema.
[21] M. Abe, G. Gopakumar, M. Hada, B. P. Das, H. Tatewaki, and D. Mukherjee, Phys. Rev. A **90**, 025001, (2014).
[22] A. S. P. Gomes, K. G. Dyall, and L. Visscher, Theor. Chem. Acc. **127**, 369 (2010); see http://dirac.chem.sdu.dk/basisarchives/dyall/#byblock.
[23] Y. Watanabe, H. Tatewaki, T. Koga, and O. Matsuoka, J. Comput. Chem. **27**, 48 (2006); see

http://ccl.scc.kyushu-u.ac.jp/˜yoshi/SFXBS/GCD.html.

[24] T. Noro, M. Sekiya, and T. Koga, Theor. Chem. Acc. **131**, 1124 (2012); see http://sapporo.center.ims.ac.jp/sapporo/Order.do.

[25] See Supplemental Material at

[26] K. P. Huber and G. Herzberg, in *MOLECULAR SPECTRA and MOLECULAR STRUCTURE IV. CONSTANTS OF DIATOMIC MOLECULES* (Van Nostrand Reinhold, New York, 1979).

[27] Huber KP, Herzberg G (2005) Constants of diatomic molecules, (data prepared by Gallagher JW andJohnson RD, III). In: Linstrom PJ, Mallard WG (eds) NIST chemistry webBook, NIST standard reference database number 69, June. National Institute of Standards and Technology, Gaithersburg, 20899 (http://webbook.nist.gov).

[28] S. Knecht, S. Fux, R. van Meer, L. Visscher, M. Reiher, and T. Saue, Theor. Chem. Acc. **129**, 631 (2011).

[29] J. Čížek, in Advances in Chemical Physics: Correlation Effects in Atoms and Molecules, volume 14 page35, edited by R. LeFebvre and C. Moser (Wiely, Chichester, 1969)

[30] I. Shavitt and R. J. Bartlett, in *Many-Body Methods in Chemistry and Physics: MBPT and Coupled-Cluster Theory* (Cambridge University, 2009).

[31] A. L. Allred and E. G. Rochow, J. lnorg. Nucl. Chem. **5**, 264 (1958).

[32] J. E. Huheey, E. A. Keiter, and R. L. Keiter, in *Inorganic Chemistry: Principles of Structure and Reactivity, 4th edition*, (HarperCollins College, New York, 1993).

[33] E. R. Meyer, J. L. Bohn and M. P. Deskevich, Phys. Rev. A **73**, 062108 (2006).

[34] P. Jönsson, X. He, C. F. Fischer, and I. P. Grant, Comput. Phys. Commun. **177**, 597 (2007).

TABLE I. Basis set information.

| Atom | Basis set |
|---|---|
| H | All of the hydrides : 7$s$, 4$p$, 3$d$ |
| F | HgF (QZ) : 13$s$, 10$p$, 3$d$, 2$f$ |
|  | Other fluorides : 13$s$, 10$p$, 4$d$, 3$f$ [17] |
| Yb | DZ : 24$s$, 19$p$, 13$d$, 8$f$, 1$g$ [17] |
|  | QZ : 35$s$, 30$p$, 19$d$, 13$f$, 5$g$, 3$h$, 2$i$ [17] |
| Hg | DZ : 24$s$, 19$p$, 12$d$, 9$f$, 1$g$ |
|  | QZ : 34$s$, 30$p$, 19$d$, 13$f$, 6$g$, 4$h$, 1$i$ |

TABLE II. Summary of the calculated results, $E_{\text{eff}}$, PDM and T1 diagnostic at the Dirac-Fock and CCSD levels.

| property | $E_{\text{eff}}$ (GV/cm) | | | | PDM (D) | | | | T1 diagnostic | |
|---|---|---|---|---|---|---|---|---|---|---|
| method | Dirac-Fock | | CCSD | | Dirac-Fock | | CCSD | | CCSD | |
| basis set | DZ | QZ | DZ | QZ | DZ | QZ | DZ | QZ | DZ | QZ |
| YbH | 21.5 | 21.8 | 25.9 | 31.3 | 2.61 | 2.62 | 2.56 | 2.93 | 0.0617 | 0.0275 |
| YbF [17] | 17.9 | 18.2 | 21.9 | 23.2 | 3.20 | 3.20 | 3.37 | 3.59 | 0.0393 | 0.0311 |
| HgH | 104.7 | 106.9 | 114.1 | 118.5 | 0.66 | 0.62 | 0.27 | 0.15 | 0.0230 | 0.0244 |
| HgH[a] [18] | - | 106.9 | - | 123.2 | - | - | - | - | - | - |
| HgF | 103.4 | 105.3 | 110.3 | 114.4 | 3.90 | 3.88 | 3.10 | 2.97 | 0.0231 | 0.0246 |

[a]Dyall.cv4z basis set was used for Hg atom, and cc-pCVQZ basis set was used for H atom.

TABLE III. MP of all the electrons in YbH, YbF, HgH and HgF. "Heavy total" denotes the sum of the MPs of the *s*, *p*, *d* and *f* orbitals of the heavy atoms and "Light total" denotes the sum of the MPs of the *s* and *p* of the light atoms. This representation is same to table IV.

|  | YbH | YbF | HgH | HgF |
|---|---|---|---|---|
| Heavy (*s*) | 10.91 | 10.87 | 11.38 | 11.12 |
| Heavy (*p*) | 24.33 | 24.15 | 24.52 | 24.31 |
| Heavy (*d*) | 20.17 | 20.16 | 29.94 | 29.96 |
| Heavy (*f*) | 14.01 | 14.01 | 14.01 | 14.05 |
| Heavy total | 69.41 | 69.18 | 79.85 | 79.44 |
| Light (*s*) | 1.58 | 4.01 | 1.12 | 3.96 |
| Light (*p*) | 0.01 | 5.81 | 0.02 | 5.59 |
| Light total | 1.59 | 9.81 | 1.14 | 9.54 |

TABLE IV. MP of SOMO electron in YbH, YbF, HgH and HgF.

|  | YbH | YbF | HgH | HgF |
|---|---|---|---|---|
| Heavy (*s*) | 0.68 | 0.86 | 0.40 | 0.71 |
| Heavy (*p*) | 0.27 | 0.13 | 0.40 | 0.18 |
| Heavy (*d*) | 0.03 | 0.02 | 0.02 | 0.04 |
| Heavy (*f*) | $2\times10^{-5}$ | $3\times10^{-4}$ | $4\times10^{-3}$ | $3\times10^{-3}$ |
| Heavy total | 0.98 | 1.01 | 0.83 | 0.93 |
| Light (*s*) | 0.02 | $-4\times10^{-3}$ | 0.17 | 0.01 |
| Light (*p*) | $6\times10^{4}$ | $-2\times10^{-3}$ | 0.01 | 0.06 |
| Light total | 0.02 | -0.01 | 0.17 | 0.07 |

TABLE V. The comparison of the AO energy differences between 6s orbital of the heavy atoms and the valence orbital of the light atoms for the four molecules, the overlap integrals and the p components of the heavy atoms of SOMO-MPs. The energies of valence orbitals of H, F, Yb and Hg ($1s$, $2p_{3/2}$, $6s$ and $6s$) were evaluated from the ground state of the neutral atoms by GRASP2K.

|  | YbH | YbF | HgH | HgF |
|---|---|---|---|---|
| energy difference (a.u.) (heavy 6s -light valence) | 0.30 | 0.54 | 0.17 | 0.41 |
| Overlap integral (heavy 6s - light valence) | 0.38 | -0.06 | 0.42 | -0.12 |
| SOMO-MP of p orbital of heavy atom | 0.27 | 0.13 | 0.40 | 0.18 |

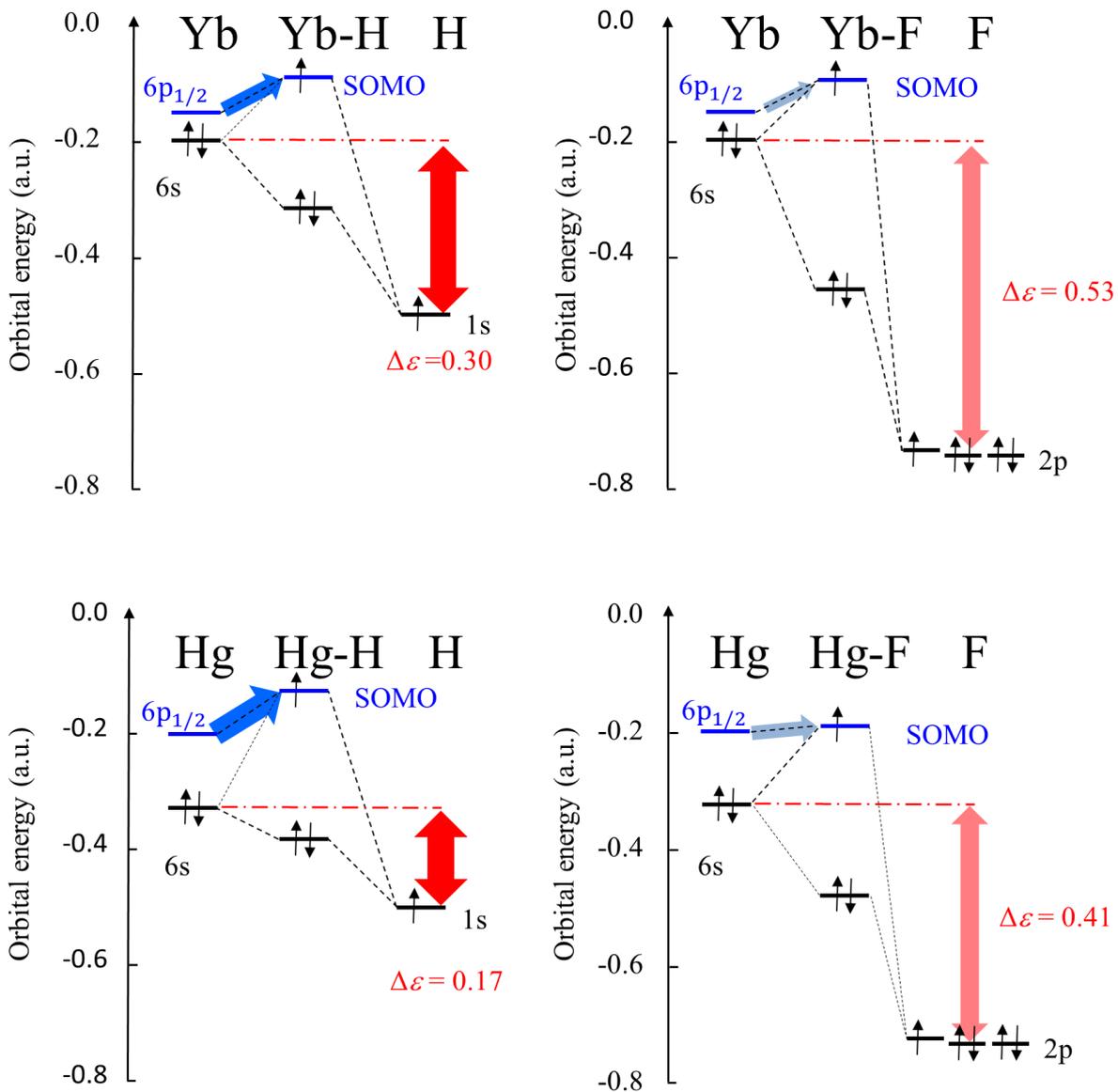

FIG.1 Energy diagram of the AO energies of H, F, Yb and Hg atom, and the SOMO and SOMO -1 energies of YbH, YbF, HgH and HgF. The energies of the valence occupied orbitals of H, F, Yb and Hg ($1s$, $2p$, $6s$ and $6s$) were evaluated from the ground states of the atoms. The $6p$ orbital energies of Yb and Hg were evaluated from the excited state of the atoms whose valence electron configurations are $6s^1 6p^1$. The atomic calculations were based on GRASP2K [26]. MO energies of the four molecules were evaluated at the DF level and QZ basis sets.